\documentclass[twocolumn]{aastex631}

\usepackage{amsmath}
\usepackage[maxfloats=256]{morefloats}
\received{October 18, 2021}
\revised{May 15, 2022}
\accepted{May 26, 2022}

\shorttitle{The Dual Radio Core Non-Detections in NGC~7674}
\shortauthors{Breiding et al.}
\graphicspath{{./}{figures/}}

\begin{document}





\title{Deep VLBI Observations Challenge Previous Evidence of a Binary Supermassive Black Hole Residing in the Seyfert Galaxy NGC~7674 }

\author[0000-0003-1317-8847]{Peter Breiding}
\affiliation{Department of Physics and Astronomy, Johns Hopkins University, Bloomberg Center, 3400 N. Charles St., Baltimore, MD 21218, USA}
\affiliation{Department of Physics and Astronomy, West Virginia University, P.O. Box 6315, Morgantown, WV 26506, USA}
\affiliation{Center for Gravitational Waves and Cosmology, West Virginia University, Chestnut Ridge Research Building, Morgantown, WV 26505, USA}

\author[0000-0003-4052-7838]{Sarah Burke-Spolaor}
\affiliation{Department of Physics and Astronomy, West Virginia University, P.O. Box 6315, Morgantown, WV 26506, USA}
\affiliation{Center for Gravitational Waves and Cosmology, West Virginia University, Chestnut Ridge Research Building, Morgantown, WV 26505, USA}
\affiliation{Canadian Institute for Advanced Research, CIFAR Azrieli Global Scholar, MaRS Centre West Tower, 661 University Ave. Suite 505, Toronto ON M5G 1M1, Canada}

\author[0000-0003-4341-0029]{Tao An}
\affiliation{Shanghai Astronomical Observatory, Key Laboratory of Radio Astronomy, CAS, 80 Nandan Road, Shanghai 200030, China}

\author{Karishma Bansal}
\affiliation{SOFIA-USRA, NASA Ames Research Center, MS 232-12, Moffett Field, CA 94035-0001, USA}

\author[0000-0002-2211-0660]{Prashanth Mohan}
\affiliation{Shanghai Astronomical Observatory, Key Laboratory of Radio Astronomy, CAS, 80 Nandan Road, Shanghai 200030, China}

\author{Gregory B. Taylor}
\affiliation{Department of Physics and Astronomy, University of New Mexico, 210 Yale Blvd NE, Albuquerque, NM 87106, USA}

\author{Yingkang Zhang}
\affiliation{Shanghai Astronomical Observatory, Key Laboratory of Radio Astronomy, CAS, 80 Nandan Road, Shanghai 200030, China}



\begin{abstract}
Previous Ku-band (15~GHz) imaging with data obtained from the Very Long Baseline Array (VLBA) had shown two compact, sub-pc components at the location of a presumed kpc-scale radio core 
in the Seyfert galaxy NGC~7674.  It was then presumed that these two unresolved and compact components were dual radio cores corresponding to two supermassive black holes (SMBHs) accreting surrounding gas and launching radio-bright relativistic jets.  
However, utilizing the original VLBA dataset used to claim the detection of a binary SMBH, in addition to later multi-epoch/multi-frequency datatsets obtained from both the VLBA and the European VLBI Network, we find no evidence to support the presence of a 
binary SMBH.  We place stringent upper limits to the flux densities of any sub-pc-scale radio cores 
which are at least an 
order of magnitude lower than the original VLBI radio-core detections, directly challenging the original binary SMBH detection claim.  With this in mind, we discuss the possible reasons for the non-detection of any VLBI radio cores in our imaging, 
the possibility of a binary SMBH still residing in NGC~7674, and the prospect of future observations shedding further light on the true nature of this active galactic nucleus.   
\end{abstract}



\section{Introduction} \label{sec:intro}

In the standard $\Lambda$CDM concordance model of cosmology, a major driver of galaxy growth and evolution is the hierarchical merging of other galaxies, subsuming each other's stars, gas, dark matter haloes, and supermassive black holes (SMBHs, M$_{\mathrm{BH}}\gtrsim 10^{6}$M$_{\odot}$) in the process \citep[e.g.,][]{kauffmann+00}.  Since most galaxies are believed to host a SMBH in their dynamical centers \citep{kormendy13}, a predicted outcome of these galaxy mergers is the formation of 
gravitationally-bound binary SMBHs \citep{begelman80}.  First, the SMBHs residing in each galaxy are thought to sink towards the nuclear environment of the post-merger remnant via dynamical friction \citep[][]{chandrasekhar43}, ultimately achieving binary separations on the order of $\sim$~10~pc within $\sim$100~Myr \citep{callegari09}.  The subsequent stage of binary evolution from $\sim$~10~pc to sub-pc separations relies on three-body interactions with stars \citep[e.g.,][]{sesana+07}, gravitational torques from a gaseous circumbinary disk \citep[e.g.,][]{escala+05}, and potentially three-body SMBH interactions \citep[e.g.,][]{hoffman+07} to harden, or shrink, the binary until gravitational radiation can efficiently remove angular momentum from the pair until their eventual coalescence. Depending on the nuclear environment of the post-merger host galaxy, this intermediate evolutionary stage between dynamical friction and gravitational radiation dominating the binary's orbital dynamics can take anywhere from 10~Myr \citep[e.g.,][]{khan15} to several tens of Gyr \citep[e.g.,][]{yu2002}, where the latter scenario is typically referred to as the ``final parsec problem'' \citep{finalpcproblem}. After surmounting the evolutionary stage pertinent to the final parsec problem, long-wavelength gravitational waves emitted by close-orbit ($\ll 0.1$~pc) binary SMBHs in the local Universe should soon be detectable by current Pulsar Timing Array experiments \citep[][]{mclaughlin+13,hobbs+13,verbiest+16,nanograv12.5} and the upcoming Laser Interferometer Space Antenna \citep[LISA,][]{lisa17}, constituting a crucial source class for these observatories \citep{gwastro-review}. 




There are several observational approaches 
other than gravitational-wave detection used to infer the presence of binary SMBHs at separations $\lesssim~10~$pc. The two main methods include the observation of quasi-periodic light curves of close-orbit ($\ll0.1$~pc) binaries \citep[e.g.,][]{graham+15,charisi+16,liu.t.+16,2016MNRAS.463.1812M,liu.t.+19,liu.t.+20} and velocity-offset broad emission lines for binaries with separations of $\sim$ 0.1 to tens of pc in nearby quasars \citep[e.g.,][]{bogdanovic+09,tsalmantza+11,Eracleous_12,ju+13,shenyue+13,runnoe+17,kelley+21}. The former involves a 
periodic modulation of the observed 
luminosity (at the orbital period of the binary), while the latter technique relies on the Doppler shifting of spectral lines emitted by gas gravitationally bound to the SMBH in the so-called broad line region (BLR, where the Doppler shifts result from the binary's orbital dynamics). However, both 
methods have a variety of more mundane explanations. In the case of periodic light curves, jet precession not induced by a binary \citep[e.g.,][]{liska+18}, a warped accretion disk \citep[e.g.,][]{hopkins+10}, helical jet morphologies \citep[e.g.,][]{conway+93}, magnetohydrodynamic (MHD) instabilities associated with the accretion flow \citep[e.g.,][]{king+13} or global oscillation of the accretion disk \citep[e.g.,][]{2013MNRAS.434.3487A,2014MNRAS.443...58W}, and stochastic variability \citep[][]{vaughan+16} are all viable alternative hypotheses to a binary SMBH. In the case of quasars with velocity-offset broad emission lines, recoiling SMBHs and BLR outflows are also plausible alternative hypotheses \citep[see e.g., ][]{breiding+21}.  

A more direct observational technique which does not suffer from the above uncertainties is the spatially-resolved imaging of both SMBHs \citep[e.g.,][]{2018RaSc...53.1211A},  given that they are both actively accreting material and shining brightly as active galactic nuclei (AGN).  This technique has revealed several dozens of dual AGN (AGN with kpc-scale separations)
, utilizing telescopes across the electromagnetic spectrum (see e.g. Table 1 from \citealt{rubinar+18} and \citealt{derosa+19} for an overview).  However, in order to resolve the pc and sub-pc spatial scales in gravitationally-bound binary SMBHs, telescopes with milliarcsecond (mas) and sub-mas resolving powers are required\footnote{Note that at a redshift of $\sim$~0.05, 1~pc corresponds to 1~mas. Even at distances of $\sim$ a few Mpc (roughly the size of the local group), sub-pc resolution can only be achieved with $\sim$~10 mas angular resolution.}. Currently, the only approach capable of achieving this type of angular resolution relies on Very Long Baseline Interferometry (VLBI) in order to synthesize the large required apertures. There have been several radio\footnote{Although see \citealt{gravity17} for an example of the promise of high-resolution optical interferometry and synthesis imaging.} VLBI studies searching for pc-scale binary SMBHs systematically in surveys \citep[e.g.,][]{sarah11,tremblay+16}, and as follow-up observations of binary SMBH candidates identified through more indirect methods \citep[e.g.,][]{kharb+17,breiding+21}. However, to date these projects have been largely unsuccessful in uncovering large numbers of bona fide binary SMBHs.  

In the context of high angular resolution radio imaging, the ``smoking gun'' signature of a binary SMBH would be the observation of two compact, and flat (or inverted) spectrum cores. The physical model used to describe these VLBI cores is the radio emission from the optical depth, $\tau~=~1$ surface at which the base of some relativistic jet become opaque to synchrotron self-absorption \citep{sokolovsky+11}, or a standing shock slightly downstream from this surface \citep{marscher+08}. In either case, the VLBI core is located several-to-tens of pc downstream of the SMBH producing the jet, and is thus a good marker for its location. At a projected separation of 7~pc, the AGN CSO~0402+379 is a striking example of a nearby ($z~=~$0.055) binary radio core confidently confirmed as a binary SMBH system \citep[][]{rodriguez06}. Using the Very Long Baseline Array (VLBA), \cite{bansal+17} have tracked the orbital motion of the cores with proper motion measurements, lending further support to the binary SMBH nature of this system.

NGC~7674 (aka MRK~533) is a nearby ($z$~=~0.03), type~II Seyfert galaxy \citep{mirabel_wilson84} home to another purported close-separation binary SMBH.  \cite{kharb+17}, referred to as K17 in the rest of the paper, claimed the detection of two inverted-spectrum radio cores in the nucleus of NGC~7674 (at a projected separation of 0.7~pc), presumably the product of a binary SMBH system in which each black hole is active and hosts a radio jet.  The host galaxy is a luminous infrared galaxy \citep[LIRG,][]{gonzalez+01}, and the brightest in a group of four interacting galaxies comprising the Hickson 96 compact galaxy group\footnote{The two largest galaxies in this group, one of which being NGC~7674, are separated by $\sim$~80~kpc (projected).  The closest galaxy companion to NGC~7674 is at a projected separation of $\sim$~30~kpc (see Figure~1 from \citealt{verdes+97} for an optical image of the compact galaxy group).  }.  It is a nearly face-on (inclination angle of $\sim$~30$^{\circ}$) spiral galaxy \citep[SBc type,][]{williams+87}, exhibiting tidal features which are likely the imprint of gravitational interactions with its neighbors \citep{verdes+97}.

Deep 2001 S-band ($\sim$~1.4~GHz) VLBA observations of NGC~7674, in combination with phased Very Large Array (VLA) and Arecibo observations, allowed for the first detection of the S (or Z)-shaped morphology of the kpc-scale radio jet \citep[][]{momjian+03}.  One hypothesized origin for S-shaped radio jet morphologies is jet precession induced by the orbital dynamics of a binary SMBH \citep{begelman80}.  However, S-shaped jets may also be caused by 
jet precession induced by a tilted accretion disk \citep[][]{sarazin+80,lu90} or the gas circulation of the interstellar medium (ISM) \citep[][]{gopal+03}.  The linear extent of the jet is  $\sim$~0.6~kpc projected on the plane of the sky, thus allowing for the classification of this source as a compact symmetric object\footnote{CSOs are young ($\lesssim~10^{5}$~yr) radio-loud AGN classified on the basis of double, symmetric, radio jets/lobes less than 1~kpc in extent \citep[][]{wilkinson+94,AnBaan12}} (CSO).  Given their small jet size, CSOs are able to interact strongly with the narrow line region (NLR) gas within the central kiloparsec of their host galaxy \citep{odea98}.  With this in mind, the observations of NLR gas outflows in NGC~7674 \citep{unger+88,shastri+06,smirnova+07} 
are naturally explained by the interaction of the jet with the surrounding medium \cite[e.g.][]{2013MNRAS.433.1161A,2019ApJ...873...11J}.      
	
In this paper, we 
use deep multi-band VLBI imaging from both the VLBA and European VLBI Network (EVN) to search for evidence of the two putative VLBI radio cores in NGC~7674 indicative of a binary SMBH. This search includes the VLBA data used to make the original dual VLBI core detection claim in addition to VLBI data from approximately a decade and a half later.  We also analyze the VLA data used in K17 to claim the presence of a single, kpc-scale radio core which corresponds to the unresolved emission from the two VLBI cores.  Finally, we assess the binary SMBH hypothesis and AGN activity of this source by putting our findings into the larger context of the other multi-wavelength observations and analyses of this source.  Throughout this paper we adopt a $\Lambda$CDM cosmology, with H$_{0}=67.74$ km s$^{-1}$ Mpc $^{-1}$, $\Omega_{\lambda}=0.69$, and $\Omega_{m}=0.31$ \citep{planck+16}.

\section{VLA Data Reduction \& Analysis} \label{sec:vla_data_analysis}

We reduced the archival Ku-band (15~GHz) radio data from project 14A-471 taken with the VLA in its A-array configuration on March 21, 2014.  The data was calibrated using the Common Astronomy Software Applications \citep[\texttt{CASA},][]{mcmullin+07} software package automatic  pipeline (version 5.4.1).  After pipeline calibration, any radio frequency interference (RFI) was removed via the automatic \texttt{rflag} routine within \texttt{CASA}. 3C~48 was the flux calibrator for our source, and several rounds of phase-only self-calibration, followed by a single round of amplitude and phase self-calibration was employed.  Finally, images were constructed from the visibility data using the CLEAN deconvolution algorithm employed by the \texttt{CASA} task \texttt{tclean}.  The final VLA image of NGC~7674 was made using Briggs weighting, a robust parameter of 0.5, and a multi-term, multi-frequency synthesis (mtmfs) deconvolution and imaging procedure with two Taylor terms. This type of analysis relies on modeling the sky intensity distribution as a Taylor polynomial, expanded about some reference frequency for the subsequent deconvolution and image reconstruction \citep{rau_cornwell_2011}.  Including the second Taylor term in the analysis allows for the measurement of spectral index information in radio maps in addition to the intensity.  We also imaged the source with a uniform weighting scheme, but the marginal improvement in angular resolution resulted in the unacceptable trade-off of a much lower signal-to-noise detection.   

\section{VLBI Observations \& Data Analysis} \label{sec:vlba_data_analysis}

 Below we describe the VLBI observations employed to create our final high-sensitivity images, using the VLBA and EVN.  All of the VLBI observations used in this paper are listed in Table~\ref{table:observations}.  Combining these disparate datasets together from both the VLBA and EVN improved the sensitivity afforded by the individual epochs.  In turn, this allowed us to test the hypothesis that dual, inverted-spectrum radio cores were present in this source, and furthermore that they were indicative of a binary SMBH (in which both black holes are active) residing in this galaxy.  All VLBI observations are phase-referenced, where we show plots of phase calibrator phase before and after calibration is applied to the high-frequency 15 and 22 GHz data obtained with the VLBA in the appendix.  Importantly, no obvious structure that would impact our phase calibration was seen in the images of any of our phase calibrators.  We show high-frequency 15 and 22 GHz images of our phase calibrators in the appendix for both the 2002 K17 detection epoch and follow-up 2018 epochs.  These phase calibrator targets include J2327+0940 and J2329+0834, which are separated from NGC~7674 by 0.9~degrees and 0.35~degrees, respectively.  In the appendix, we also show the \textit{(u,v)} plane coverage for each individual 15/22 GHz VLBI epoch, in addition to the \textit{(u,v)} coverage corresponding to the combined 15 and 22 GHz high-frequency datasets.  
 
 
 \def\arraystretch{1}%
 \setlength\tabcolsep{3pt}
 \begin{deluxetable}{lccr}[]	\tablecaption{\label{table:observations} VLBI Observing Sessions}
 	\tablecolumns{4}
 	\tablewidth{0pt}
 	\tablehead{
 		&Project&Observation&Observing\\
 		Observatory&Code&Date (UTC)&Band\\
 		&&(YYYY\slash MM\slash DD)&
 		}
 	\startdata
	VLBA&BV045&2002/08/28&S\\
	VLBA&BV045&2002/08/28&C\\
	VLBA&BV045&2002/08/28&X\\
	EVN&EA059&2018/06/05&X\\
	VLBA&BV045&2002/08/28&Ku\\
	VLBA&BK212&2018/03/31&Ku\\
	VLBA&BT143&2018/11/29&Ku\\
	VLBA&BK212&2018/03/19&K\\
	VLBA&BK212&2018/04/12&K\\
	VLBA&BT143&2018/11/29&K\\
	\enddata
 	\tablecomments{Observing dates mark the start of an observing session if the observations spill over into subsequent days. }
 \end{deluxetable} 
  
\subsection{VLBI Observations \& Calibration}
\label{sec:vlba}


\subsubsection{VLBA Project BV045}

The original VLBI dataset used to assert the existence of a binary SMBH in NGC~7674 was obtained from phase-referenced VLBA Ku-band (15~GHz) observations (project ID BV045) taken in 2002.  Associated with this project were observations at S, C, and X bands (2.3, 5, 8~GHz, and 15~GHz).  The experimental setup included only a single polarization recording capability, 4.2~s integration times, and a total bandwidth of 32~MHz, split into four spectral windows (with 16 channels per spectral window).  These observations included nine VLBA antennas (excluding the Brewster station), and the cycle time used  for the ``nodding'' mode observations (phase-target-phase scans) was $\sim$~11~min ($\sim$~4 min on the phase calibrator J2329+0834, $\sim$~7 min on NGC~7674).  The quasar 3C~84 was used as both the fringe finder and bandpass calibrator for each set of observations (i.e., each observing band) in BV045.

\subsubsection{VLBA Projects BK212 \& BT143}


VLBA projects BK212 and BT143 involved follow-up observations of NGC~7674 at Ku and K bands (15~GHz and 22~GHz, respectively) in 2018.  The experimental setups for both BK212 and BT143 included full polarization capabilities, 2~s integration times, and a total bandwidth of 256~MHz, split into eight spectral windows (where BK212 had 16 channels per spectral window and BT143 had 64 channels per spectral window).  BK212 included at least nine VLBA antennas, and BT143 had at least eight VLBA antennas.  The cycle time used  for the ``nodding'' mode observations (phase-target-phase scans) was $\sim$~7~min  for BK212 ($\sim$~4 min on the phase calibrator J2329+0834, $\sim$~3 min on NGC~7674).  For BT143, the cycle time was $\sim$~3~min ($\sim$~2 min on the phase calibrator J2327+0940, $\sim$~1 min on NGC~7674).  The quasars 3C~454.3 and 3C~345 were used as the fringe finder/bandpass calibrators for BK212 and BT143, respectively.



\subsubsection{VLBA calibration} 


We calibrated all of the VLBA visibility data in AIPS \citep{moorsel+96} using the standard calibration procedures applied in the pipeline task \texttt{VLBARUN} for continuum imaging, where appropriate reference antennas were chosen.  Log-based flagging was performed prior to calibration.

\subsubsection{EVN Observations \& Calibration}
\label{sec:evn}
The target source was also observed by EVN on June 6th, 2018, (project code EA059A) with the aim of probing the dual cores.  The observations were conducted at 8.4 GHz with a total observing time of 10 hours. The session was recorded at 1024 Mbps rate (16 MHz $\times$ 8 subbands, 2-bit sampling, dual polarization). 12 stations participated in session A, which are Ef (Effelsberg, Germany), Wb (Westerbork, The Netherlands), Mc (Medicina, Italy), Nt (Noto, Italy), O6 (Onsala, Sweden), Ys (Yebes, Spain), Hh (Hartebeesthoek, South Africa), Sv (Svetloe, Russia), Zc (Zelenchukskaya, Russia), Bd (Badary, Russia), Ir (Irbene, Latvia), T6 (Tianma65, China).

During the observations, the phase-reference observing mode was employed due to the weakness of the target source.
J2327+0940 was chosen as the phase-reference calibrator and several scans on J2329+0834 were observed for finding fringes.
To reduce the impact from the dynamic troposphere, we used a shorter nodding cycle at shorter wavelengths.  We used a cycle time of 'cal(30s)-tar(135s)-cal(45s)'.
Due to some operational problems and bad fringe solutions at a few stations, some antennas did not record well during specific time periods. These bad data were deleted.
After the observations were completed, the data from each station were transported to JIVE at Dwingeloo, the Netherlands for correlation. The correlated visibility data were then downloaded to the China SKA Regional Centre computing clusters \citep{2019NatAs...3.1030A} for further analysis and processing.
We calibrated the visibility data using AIPS following a standard procedure used for phase-reference EVN observations of weak radio sources \citep[e.g.,][]{2020ApJ...888L..24M,2021arXiv210607169S}. 

\subsection{Data Combination \& Imaging}
\label{sec:imaging}

 \def\arraystretch{0.85}%
\setlength\tabcolsep{3pt}
\begin{deluxetable}{lccccr}[]	\tablecaption{\label{table:imaging} VLBI Image Properties}
	\tablecolumns{6}
	\tablewidth{0pt}
	\tablehead{
		Observing&Central&Image&&&\\
		Band&Frequency&Noise Level&$\mathrm{B_{maj}}$&$\mathrm{B_{min}}$&P.A\\
		&(GHz)&($\mathrm{\mu}$Jy beam$^{-1}$)&(mas)&(mas)&($^{\circ}$)
	}
	\startdata
	S&2.27&126&11.5&4.21&$-$14.5\\
	C&5.00&90.0&4.40&1.74&$-$10.0\\
	X&8.41&86.9&2.77&1.23&$-$7.40\\
	Ku&15.3&40.0&1.35&0.54&$-$6.76\\
	K&22.2&45.2&1.59&0.43&$-$14.7\\
	\enddata
\end{deluxetable} 

After calibration, we used \texttt{CASA} for further flagging of RFI.  Subsequently, we combined all of the visibility data from the different observatories and epochs common to a given observing band using the \texttt{CASA} task \texttt{concat}, and then cleaned/imaged that data with the \texttt{CASA} task \texttt{tclean}. We used a natural weighting scheme, as this has the highest sensitivity for point sources (which is the expectation for the two unresolved, compact cores), with cell sizes of $\sim$~4~pixels per restoring beam.  We used the \texttt{hogbom} deconvolution algorithm, and an interactive cleaning procedure with CLEAN masks ultimately created for components C and W during the cleaning process.  At no point did any significant residuals suggest the need for cleaning any image components near the purported location of the binary SMBH.  All cleaning had 3$\sigma$, rms-based threshold minor cycle stopping points.  

We also made wide-field images to search for any VLBI cores on scales up to $\sim$~1\,arcsec from the phase center.  This was accomplished with the \texttt{wproject} gridding algorithm \citep[][which corrects for the effect of non-coplanar baselines]{cornwell+08}, with the use of \texttt{wprojplanes=-1} in \texttt{CASA}.  This choice automatically determines the number of planes to use based upon the data and image size\footnote{\texttt{CASA} determines this number based upon the following formula: $\mathrm{N_{wprojplanes}=0.5\times\frac{W_{max}}{\lambda}\times\frac{imsize}{(radians)}}$.  Here,  imsize is the image size, $\lambda$ is the wavelength, and $\mathrm{W_{max}}$ is the maximum w in the uvw data (i.e., physical extent of the visibility data that is orthogonal to the image plane).}.

\section{Results} \label{sec:results}

\begin{figure}[]
	\begin{center}
		\includegraphics[scale=0.47]{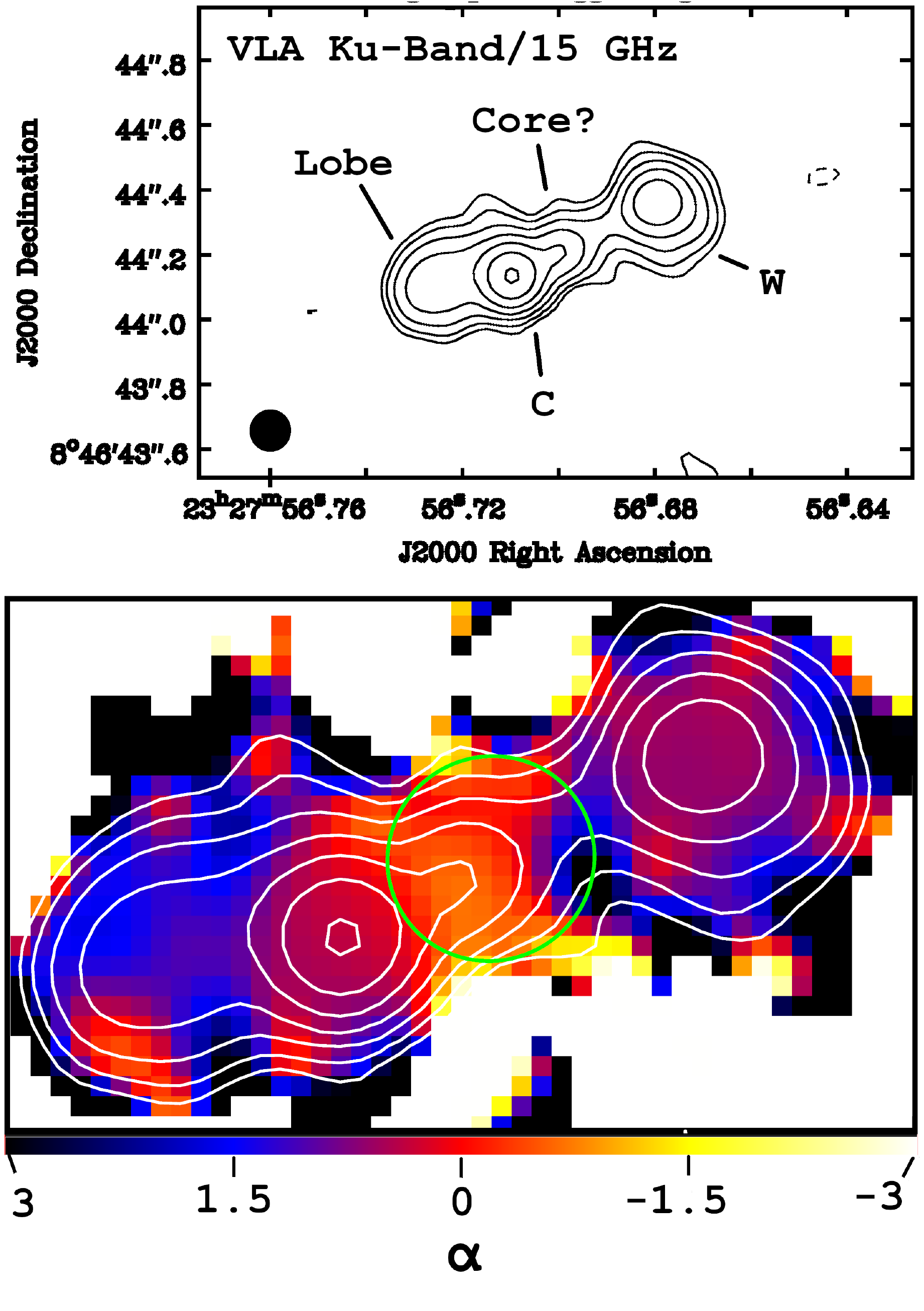}
	\end{center}
	\caption{ In the top panel we show the VLA Ku-band (15~GHz) radio image of NGC~7674 as a contour map, with a base level of 3$\sigma$ and spaced by factors of two thereafter (RMS noise level is 0.7~$\mathrm{\mu Jy~beam^{-1}}$).  Negative contours are shown as dashed lines.  Components labeled C and W correspond to the hot spots, following the naming convention used in previous studies.  We also label the feature claimed to correspond to the radio core in K17, in addition to lobe plasma east of component C.  The synthesized beam is shown in the bottom left as a filled ellipse.  The bottom panel shows the spectral index map, with white contours corresponding to the image contours presented in the top panel.  The green ellipse shows the region used to extract the spectral index information for the region corresponding to the purported radio core.}
	\label{fig:vla_figure}
\end{figure}

\begin{figure*}[]
	\begin{center}
		\includegraphics[scale=0.7, angle=90]{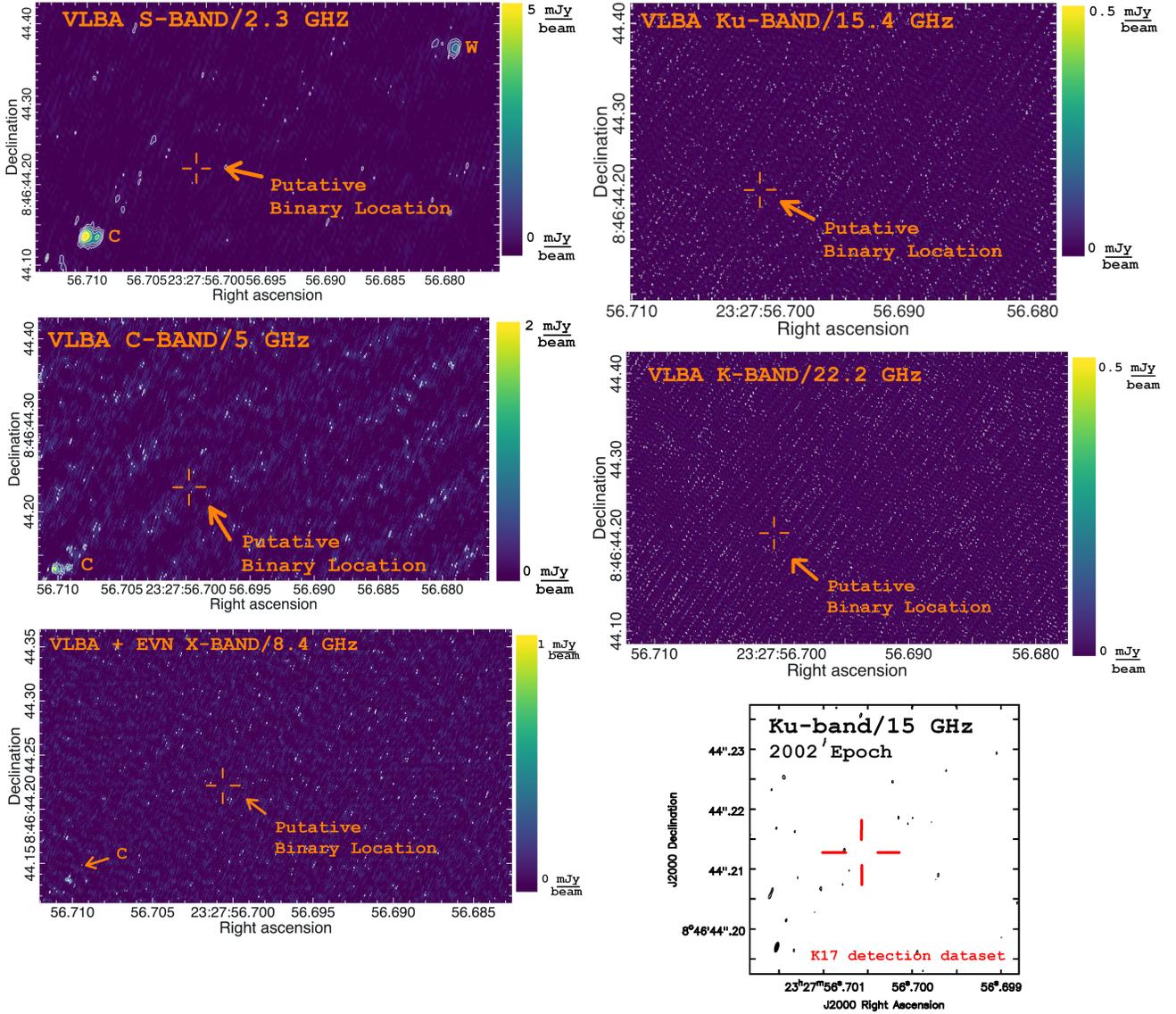}
	\end{center}
	\caption{ High-sensitivity VLBI images of NGC~7674 at S (top left panel, 2.3~GHz), C (middle left panel,5~GHz), X (bottom left panel, 8.4~GHz), Ku (top right panel, 15.4~GHz), and K (middle right panel, 22.2~GHz) bands, using the datasets listed in Table~\ref{table:observations}.  The bottom-right panel shows the contour-only image of the Ku-band (15~GHz) K17 binary detection dataset.  We show image contours for all images at a base level of 3$\sigma$,  spaced by factors of two thereafter.  The synthesized beam shapes are shown in the bottom left of the images as filled gray ellipses with a cross inside, except the bottom right contour plot where it is a filled black ellipse.  The left-panel images show the eastern and western hot spots, labeled C and W, respectively, are detected in our S/C/X-band imaging, as also shown in K17 (although component W was not detected in our C or X-band imaging).  The center position bewteen the dual cores reported by K17 is marked with cross hairs in each image, where both the dual core separation and errors of the purported core locations are on the order of the beam sizes shown for our images.  As is evident, these radio cores are not detected in any of these images.}
	\label{fig:vlbi_images}
\end{figure*}

\subsection{VLA imaging and spectral analysis}

In Figure~\ref{fig:vla_figure} we show the 15~GHz radio image of NGC~7674, synthesized from 2014 A-configuration VLA observations and label the jet features following the conventions in previous studies of the source.  The feature most relevant to this analysis is between the eastern/western hot spot components (C/W), and is identified as the radio core in K17.  We show in the bottom panel of Figure~\ref{fig:vla_figure} the in-band spectral index, $\alpha$, map of NGC~7674.  For this study, we assumed a spectral index defined as $F_{\nu}\propto\nu^{-\alpha}$, where $F_{\nu}$ is the flux density, and $\nu$ is the observing frequency.  Using an elliptical extraction region (with the same shape as the synthesized beam) across the feature identified as the radio core, we measured the mean spectral index from this region to be $-0.66~\pm~0.52$.  This index is consistent with the claimed inverted-spectrum index found in K17.  
\subsection{Non-detections of VLBI radio cores}

In Table~\ref{table:imaging} we give the central observing frequencies, RMS noise levels, major and minor axes ($\mathrm{B_{maj}}$ and $\mathrm{B_{min}}$, respectively) of the synthesized beam full width at half maximums (FWHMs), and beam position angles (P.A.) for the final high-sensitivity images used in our study (refer to Table~\ref{table:observations} for the list of datasets that went into creating each image).  In Figure~\ref{fig:vlbi_images} we show the S, C, and X-band images in which we can report detections of the hot spot components (C and W).  As shown in K17 and found again in our imaging, these hot spots have a fairly steep spectrum indicative of aged, optically thin plasma emitting synchrotron radiation. In Figure~\ref{fig:vlbi_images} we also show the Ku-band (15~GHz) and K-band (22~GHz) synthesized VLBI images for the combined Ku and K-band data described in section~\ref{sec:vlba}, in addition to an image of the original Ku-band K17 detection dataset.  We mark the locations of the dual VLBI core detections from K17 with cross hairs (taking the center of the cross hair to be the center position between the dual core positions), where we estimate the error on the core positions from K17 to be no greater than $\sim~0.7$~mas\footnote{For this estimate we note that the phase calibrator position error is negligible (0.15~mas), and the error associated with the phase-referencing technique is expected to be on the order of $\sim$~0.1~mas \citep{pradel+06}.  Thus, these errors would add negligible contributions, considering the beam size of the tentative detection reported in K17 is 0.7~$\times$~0.7~mas.}, a size which is roughly on the order of the beam widths shown in Figure~\ref{fig:vlbi_images}.

\begin{figure}[]
	\begin{center}
		\includegraphics[scale=0.5]{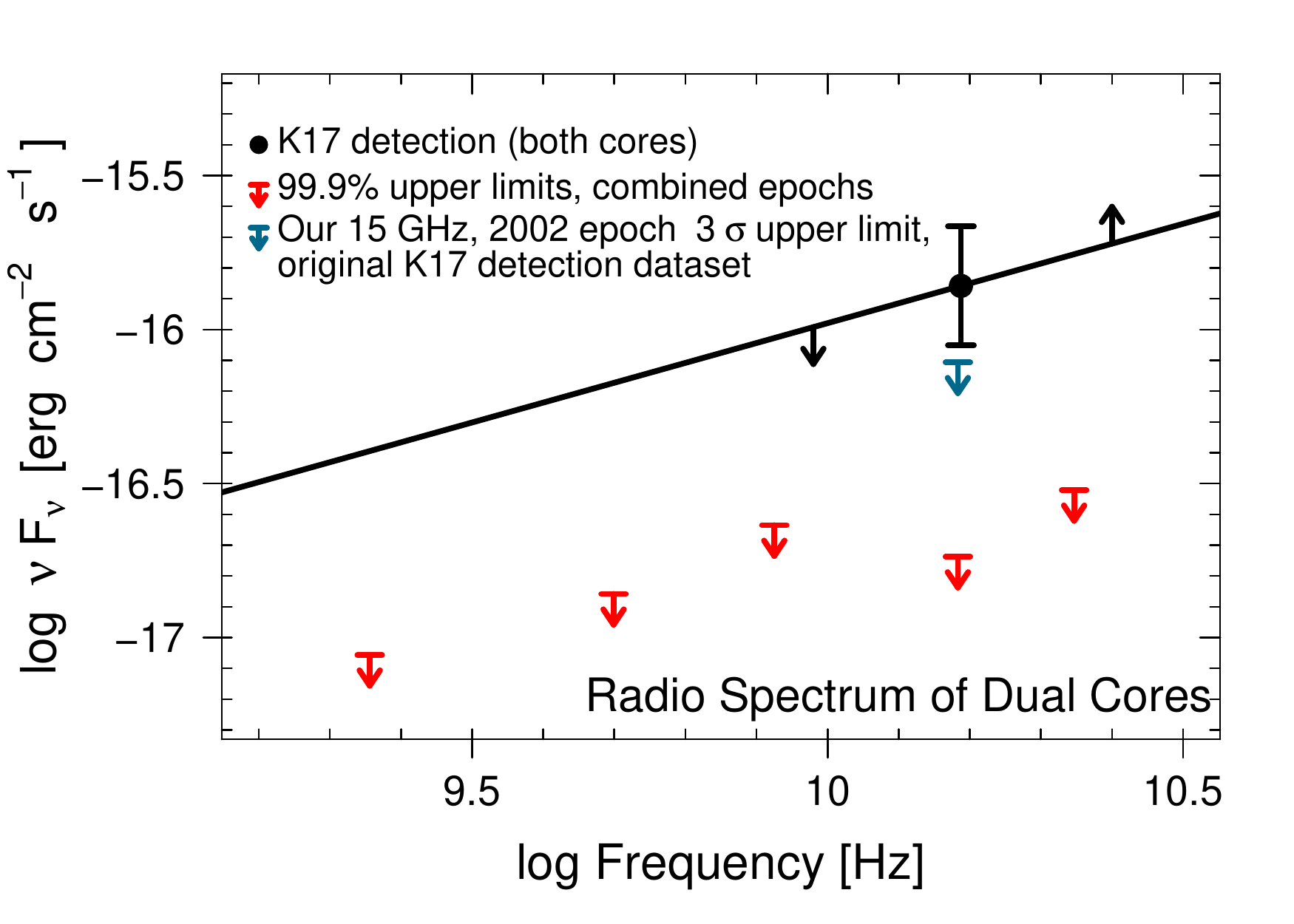}
	\end{center}
	\caption{The radio flux densities of both VLBI radio cores detected at 15~GHz by K17 are shown in black with associated 1$\sigma$ error bars.  The solid black line with arrows shows the upper limit on the radio spectral index reported by K17 for the western VLBI core based upon a X-band non-detection.  The red arrows show the 3$\sigma$ upper limits on the flux densities for either putative VLBI core based upon the non-detections of our high-sensitivity data combinations at S, C, X, Ku, and K bands.  As is apparent from the radio spectrum, the non-detections at Ku and K bands are especially problematic for the original detection claim, given the limit on spectral index and assumed inverted spectrum for the putative VLBI radio cores.}
	\label{fig:radio_spectrum}
\end{figure}

In Figure~\ref{fig:radio_spectrum}, we plot the flux density reported for both VLBI cores from K17 (0.9~$\pm$~0.4~mJy).  We also show a line (with arrows) representing the 4$\sigma$ upper limit reported by K17 on the spectral index for the western VLBI core, i.e., $\alpha~<0.38$ (where this limit is less restrictive than that of the eastern VLBI core, $\alpha~<0.33$).  In red arrows, we show our 3$\sigma$ upper limits on the flux density of either VLBI core using the images displayed in Figure~\ref{fig:vlbi_images}, where it is clear that neither core is detected in our imaging.  We also plot as a blue arrow the resulting 3$\sigma$ upper limit from our reanalysis of the original K17 Ku-band (15~GHz)  ``binary detection'' dataset.  Given the spectral index limit, the K-band non-detections are especially problematic for the hypothesis of dual, inverted-spectrum VLBI radio cores with $\sim$~1~mJy flux densities residing in the heart of NGC~7674. 



\section{Discussion \& Concluding Remarks} \label{sec:discussion}

Our upper limits obtained from our VLBI core non-detections at 15 and 22~GHz are $\sim$~an order of magnitude 
below the flux densities reported by K17 for the purported dual, inverted-spectrum VLBI cores.  Thus, we consider the possibility of the K17 detections to be false-alarm detections below (where the  original detection confidence levels were reported to be between 4$\sigma$ and 5$\sigma$).  In support of this idea is the fact that the 2002 dataset analyzed by K17 for the original dual radio core detections was also used in our analysis (BV045, comprised of S, C, X, and Ku-band VLBA observations).  While we report the high signal-to-noise detection of the hot spot components (just as in K17, see Figure~\ref{fig:vlbi_images}), there is no evidence for the putative binary.  If there was some systematic error in our analysis, we would expect that this would also affect our detection of the hot spot features (which was not the case).  Furthermore, the Ku-band and K-band observations from 2018 yield non-detection upper limits well below the expected flux densities for these cores given their supposed inverted spectrum.  One way out of this conclusion is to presume flux density variability of the cores could allow them to remain undetected in the recent 2018 K-band and Ku-band imaging.  The implied large-amplitude variations are incompatible with typical intra/inter-day stochastic variability typically observed in radio light curves of flat spectrum AGN cores and usually ascribed to interstellar scintillation \citep[e.g.,][]{lovell+08,koay+18}.  Furthermore, we note that the luminosity of the core is relatively weak compared to the hot spot/lobe plasma, implying this source is not consistent with a highly variable blazar.    
Rather, this long-term variability would require both cores to dim by more than an order of magnitude in luminosity in $\sim$ 16 years.  This hypothesis is somewhat compromised considering the feature detected in the 2014 VLA image (Figure~\ref{fig:vla_figure}) between the eastern and western hot spot components (C and W, respectively), claimed by K17 to be the unresolved emission from the dual VLBI cores.  Together these hypotheses imply that the emission from the binary AGN held a relatively constant 15~GHz luminosity from 2002 to 2014, and then dimmed by more than an order of magnitude from 2014 to 2018.  
This scenario can be tested with future, deep Ku-band (and higher frequency) VLA A-configuration observations, combined with simultaneous VLBI imaging (possibly in phased VLA and VLBA observations, as in \citealt{momjian+03}).  If the feature identified as the kpc-scale core by K17 is not detected in such an experimental setup, then this lends credence to the idea that the dual radio jets in NGC~7674 have ``turned off''.  This type of scenario could be explained by long-term ``geometric'' variability induced by e.g. jet precession (potentially due to a binary), and electromagnetic radiation which is progressively relativistically beamed \citep{blandford79} in a direction  further from our line of sight (as the approaching jets swing away from us), subsequently yielding lower observer-frame radio luminosities \citep[e.g.,][]{bach+06}. In principle, the jet precession induced by a binary SMBH system can result from two scenarios. In the first \cite[disk precession, e.g. ][]{1997ApJ...478..527K,2007ApJ...671.1272L,2021ApJ...908..178N}, the accretion disk surrounding the primary SMBH (hosting the jet) precesses due to torque action owing to a mis-alignment between the orbital plane of the binary system and the disk plane. In the second \cite[geodetic precession, e.g. ][]{begelman80,2004ApJ...615L...5R,2007ApJ...671.1272L}, the direction of spin of the primary SMBH may be mis-aligned with the total angular momentum of the binary system, thus setting up a precession of the spin axis and hence, the jet hosted by the primary SMBH-accretion disk system. The Keplerian angular frequency, and those in the cases of disk precession \citep{1997ApJ...478..527K} and geodetic precession \citep{2004ApJ...615L...5R} are
\begin{align}
\Omega_K &= \left(\frac{G M_\bullet (1+q)}{d^3}\right)^{1/2} \\ 
\Omega_D &= \frac{3 q \cos \theta_0}{4 (1+q)^{1/2}} \left(\frac{r_D}{d}\right)^{3/2} \Omega_K \\ 
\Omega_G &= q M_\bullet \left(\frac{4+3 q}{1+q}\right) \frac{G \Omega_K}{2 d c^2} ,
\end{align}
where $G$ is the gravitational constant, $M_\bullet$ is the mass of the primary SMBH, $q \leq 1$ is the mass ratio between the secondary companion and the primary SMBH, $d$ is the binary separation, $\theta_0$ is the angle of inclination between the accretion disk and the binary orbital plane, and $r_D$ is the extent of the accretion disk. We use the choices $M \approx 10^7~M_\odot$ \citep{2002ApJ...579..530W,kharb+17}, $d = 0.35$ pc \citep{kharb+17}, $\theta_0 = 20^\circ$ \citep{1997ApJ...478..527K}, and a disk extent \citep{2016MNRAS.456L.109K,2018MNRAS.473.3638G}
\begin{equation}
r_D = (5.45 \times 10^{16}~{\rm cm})~\dot{m}^{-8/27} (M/10^7 M_\odot)^{1/27},
\end{equation}
where $\dot{m}$ is the ratio of the mass accretion rate scaled in terms of the Eddington rate and is set to 0.1, appropriate for accretion disks in Seyfert galaxies \cite[e.g.][]{2014ApJ...791...74M}. With these assumptions, we obtain minimum precession periods for the case $q = 1$ (equal mass binary), with $P_D = 2 \pi/\Omega_D \geq 2.6 \times 10^5$ yr. and $P_G = 2 \pi/\Omega_G \geq 1.8 \times 10^9$ yr. These correspond to angular precessions of $\sim$ 4.9 arcsec yr$^{-1}$ and $\sim$ 0.2 mas yr$^{-1}$, respectively. A maximal value of $\sim$ 0.02$^\circ$ may thus be inferred for the case of disk precession, considering a luminosity dimming timescale of $\sim$ 16 yr. For a typical jetted AGN with an inclination angle of a few degrees \cite[e.g.][]{2019ApJ...874...43L}, this precession is two orders of magnitude smaller; changes due to relativistic beaming effects may not be discernible within the observation window. Thus, a binary SMBH enabled jet precession is unlikely to result in a significant dimming of the luminosity and we place a low credence on this possibility for our radio core non-detections.

Alternatively, it is possible mass loading mediated by jet-ISM feedback, or some other jet disruption/diminishment mechanism (potentially due to variable accretion) is responsible for the radio core non-detections we observed in the 2018 epoch and the short observed duty cycle\footnote{\cite{momjian+03} estimate the age of the AGN in NGC~7674 to be $\sim$~a few Myr, based upon the time necessary to inflate its lobes into the ambient medium.  See \citet{jurlin+20} for a longer discussion of radio-jet duty cycles in AGN.} for this AGN \citep[][]{hardcastle_croston_14,shabala+20}.   

However, a phased VLA/VLBA setup also allows for a sufficient mixture between small and large-distance baselines to determine the actual scale (i.e., between sub-pc to kpc scales) and morphology of this emission component in the event that we can detect this feature.  The future next-generation Very Large Array \citep[ngVLA,][]{murphy} would naturally allow for this mixture between small and large-distance baseline lengths to allow for sensitivities to features of varying size in this radio-loud AGN from sub-pc to kpc scales.  Our analysis of the VLA in-band spectral index of the purported core feature is consistent with its flat or inverted spectrum, and thus its interpretation as a synchrotron self-absorbed radio core.  However, as suggested by \cite{momjian+03}, the inverted radio spectrum of this jet region could be a consequence of free-free absorption \citep[FFA, see models by e.g. ][]{bicknell+97,begelman99}.  The FFA hypothesis for the inverted spectrum is strengthened by the work of \cite{gandhi+17}, who find large hydrogen column densities associated with the Compton-thick X-ray AGN in NGC~7674 from recent \textit{NuStar} observations, in addition to ionized Fe-line emission (together these findings suggest a high density of hydrogen gas and a high intensity of ionization continuum radiation in the vicinity of the central AGN).  Considering both hot spots and lobes detected from this young radio-loud AGN (see Figure~1 from \citealt{momjian+03} for a clear detection of lobes on either side of both hot spots and depiction of the overall S-shape), the naive expectation would be a relatively misaligned radio jet.  This would imply the core emission could be relativistically beamed out of our line of sight, leading to the expectation of very weak emission in the direction of the observer.   
	
If, as we 
suspect, the dual radio core detections reported by K17 correspond to spurious noise features, then presumably there is still at least one radio core of lower luminosity which has yet to be detected.  In this vein, even-deeper high-sensitivity VLBI observations could help detect this feature.  One property of this system which still suggests a binary SMBH may be present is the overall S-shape of the radio jets.  However, we note that the ``curve'' portion of the S-shape in this jet is made by the lobe plasma, after it leaves the working surfaces corresponding to the hot spots where the jet is assumed to be pushing against the ambient medium.  Since the lobe plasma does not have the momentum or kinetic power carried by the twin radio jets, it is much more likely the lobes could be pushed into the observed anti-aligned curves by the circulating ISM gas pressure (see \citealt{gopal+03} for a discussion of this model in generating S-shaped jets).

%

\begin{acknowledgments}
	We thank the anonymous referee for the many useful and constructive comments which ultimately helped improve the quality of the manuscript.  TA thanks the grant support by the Youth Innovation Promotion Association of CAS. The EVN data analysis has used the China SKA Regional Center prototype \citep{2019NatAs...3.1030A} funded by the National Key R\&D Programme of China (under grant number  2018YFA0404603) and Chinese Academy of Sciences (under grant number 114231KYSB20170003). YZ thanks Jun Yang for helping with VLBI data processing.
\end{acknowledgments}

\vspace{5mm}
\facilities{VLBA, EVN, VLA}


\software{astropy \citep{2013A&A...558A..33A,2018AJ....156..123A}, CASA, AIPS}




\appendix

Below we show the \textit{(u,v)} plane coverage for the high-frequency Ku (15~GHz) and K-band (22~GHz) VLBI visibilities used in this study, along with plots demonstrating the proper phase calibration of these visbilities and high-frequency images of our phase calibrators.  The purpose of these figures is to demonstrate the improvements in \textit{(u,v)} plane coverage upon data combination and the fact that the non-detection of any VLBI radio cores in this work is not a result of improper calibration.  

\begin{figure}[]
	\begin{center}
		\includegraphics[scale=0.75]{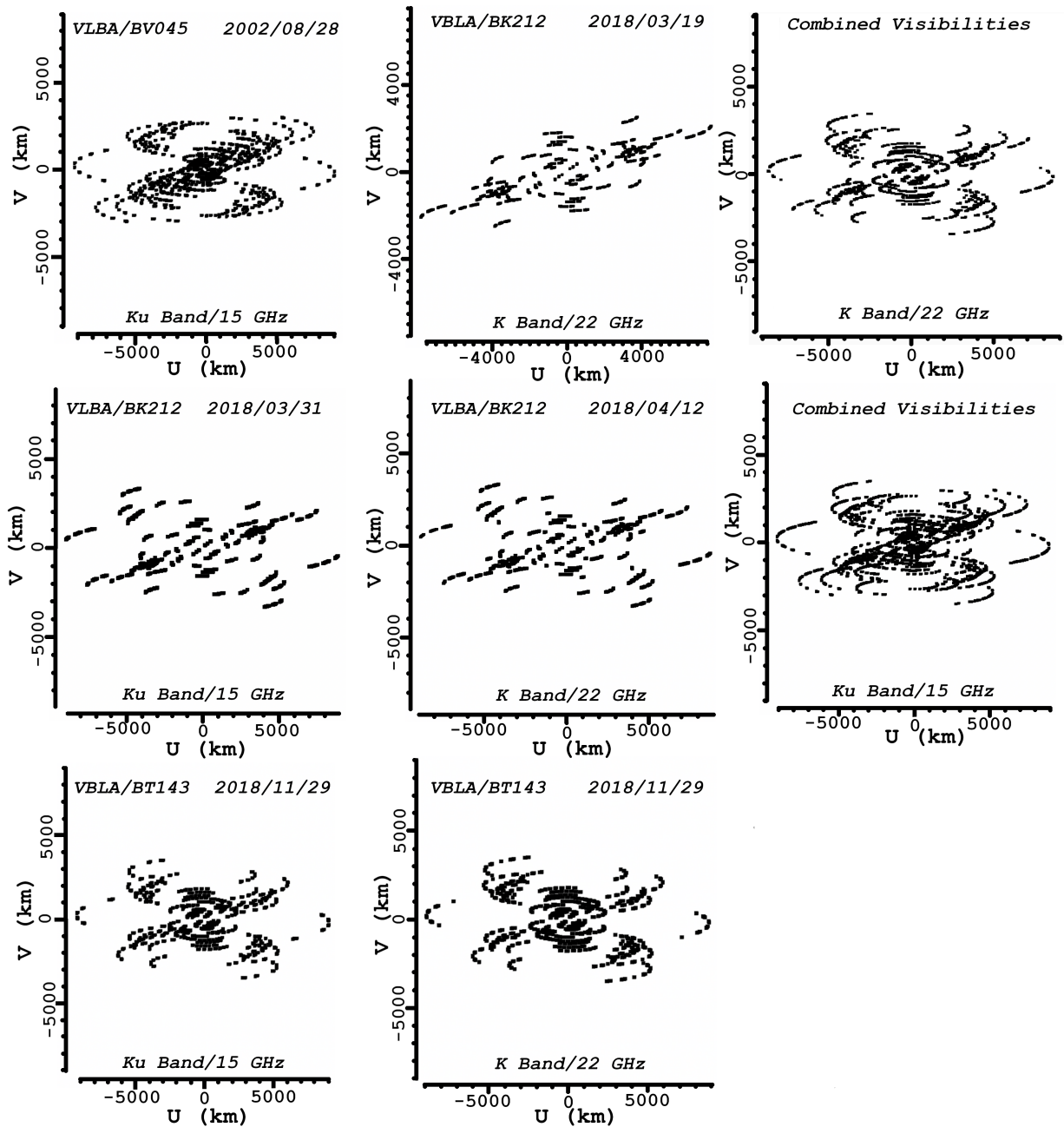}
	\end{center}
	\caption{\textit{(U,V)} plane coverage for all of the individual epoch, high-frequency Ku and K-band (15/22 GHz) VLBI vsibility data used in this study.  The VLBI observatory along with project code is given in the top left of each plot, and the date of the observation is given in the top right as YYYY/MM/DD. The frequency/band of the observation is also labeled in the bottom center of each plot.}
	\label{fig:uvplane}
\end{figure}

\begin{figure}[]
	\begin{center}
		\includegraphics[scale=0.4]{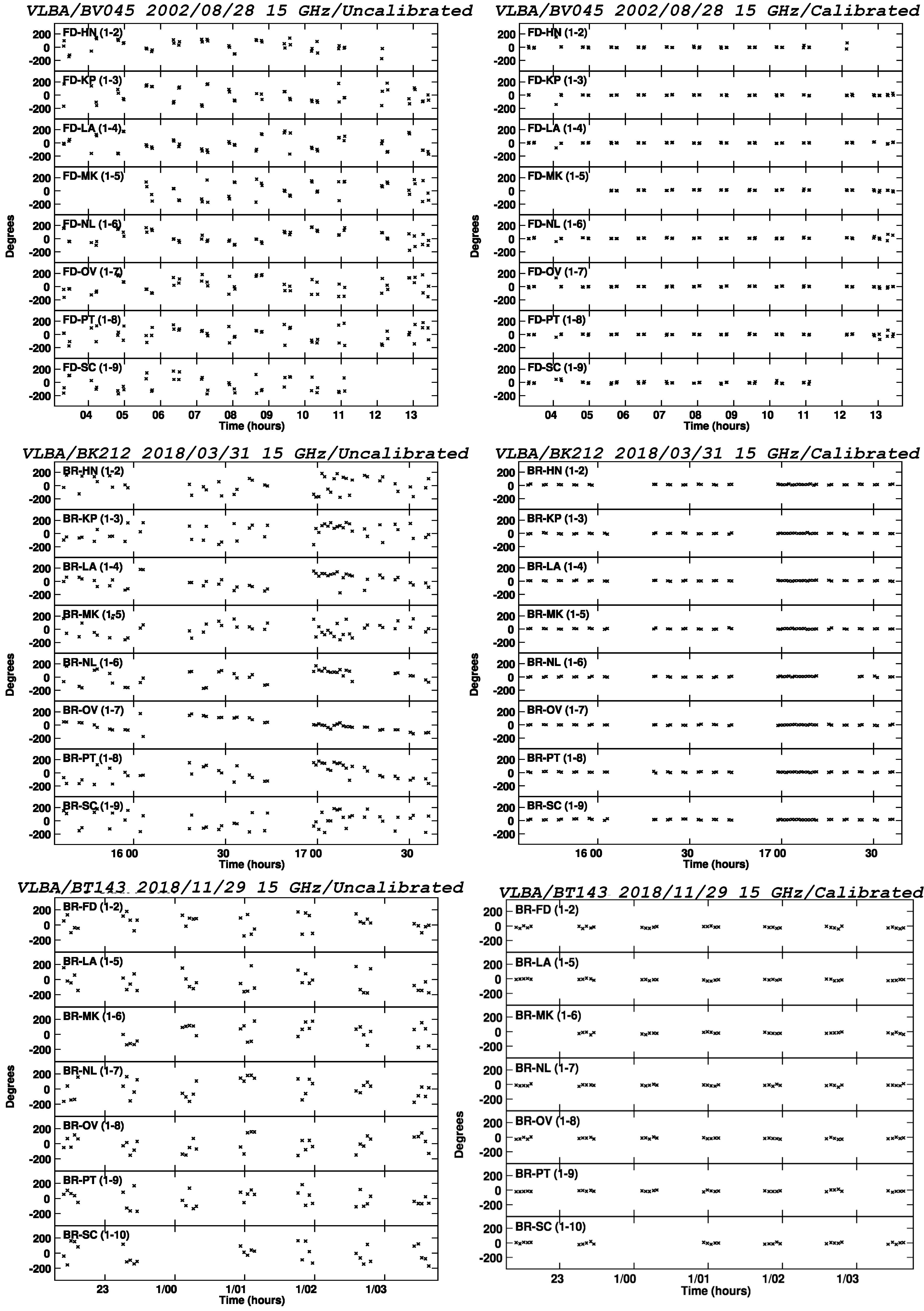}
	\end{center}
	\caption{Ku-band (15~GHz) phase of each observation's phase calibrator before (left plots) and after (right plots) phase calibration is applied for a selection of representative VLBA baselines.  The data are frequency averaged and we use a phase-solution interval ranging from 15~s to the scan length for calibration depending upon which resulted in the best phase calibration.  Observatory and project code are given above each plot, along with date of observation and the designation ``Uncalibrated'' or ``Calibrated'' representing phase data before or after calibration, respectively.}
	\label{fig:u_phase}
\end{figure}

\begin{figure}[]
	\begin{center}
		\includegraphics[scale=0.4]{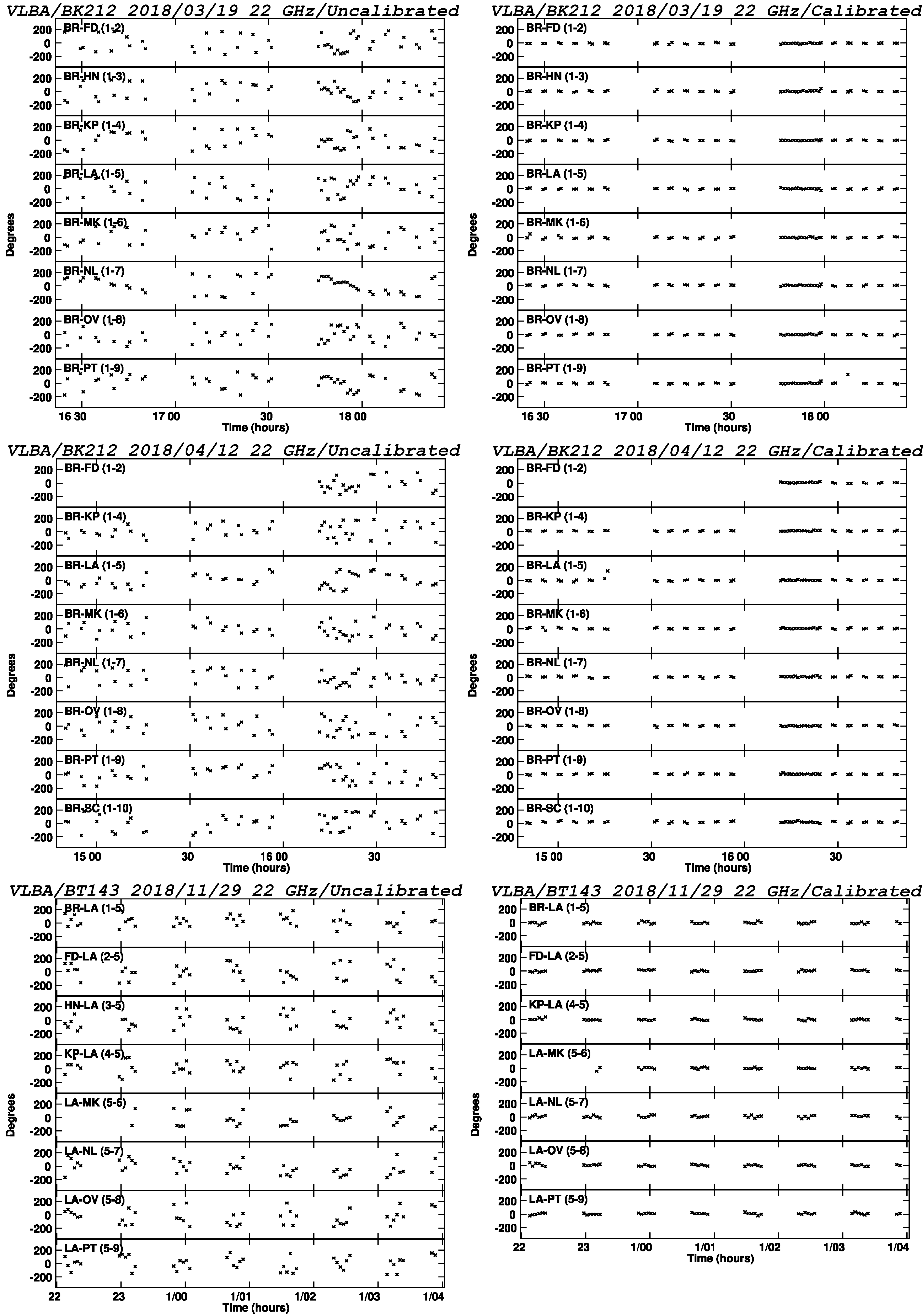}
	\end{center}
	\caption{A continuation of Figure~\ref{fig:u_phase} for the K-band, 22~GHz VLBA data.}
	\label{fig:k_phase}
\end{figure}

\begin{figure}[]
	\begin{center}
		\includegraphics[scale=0.3]{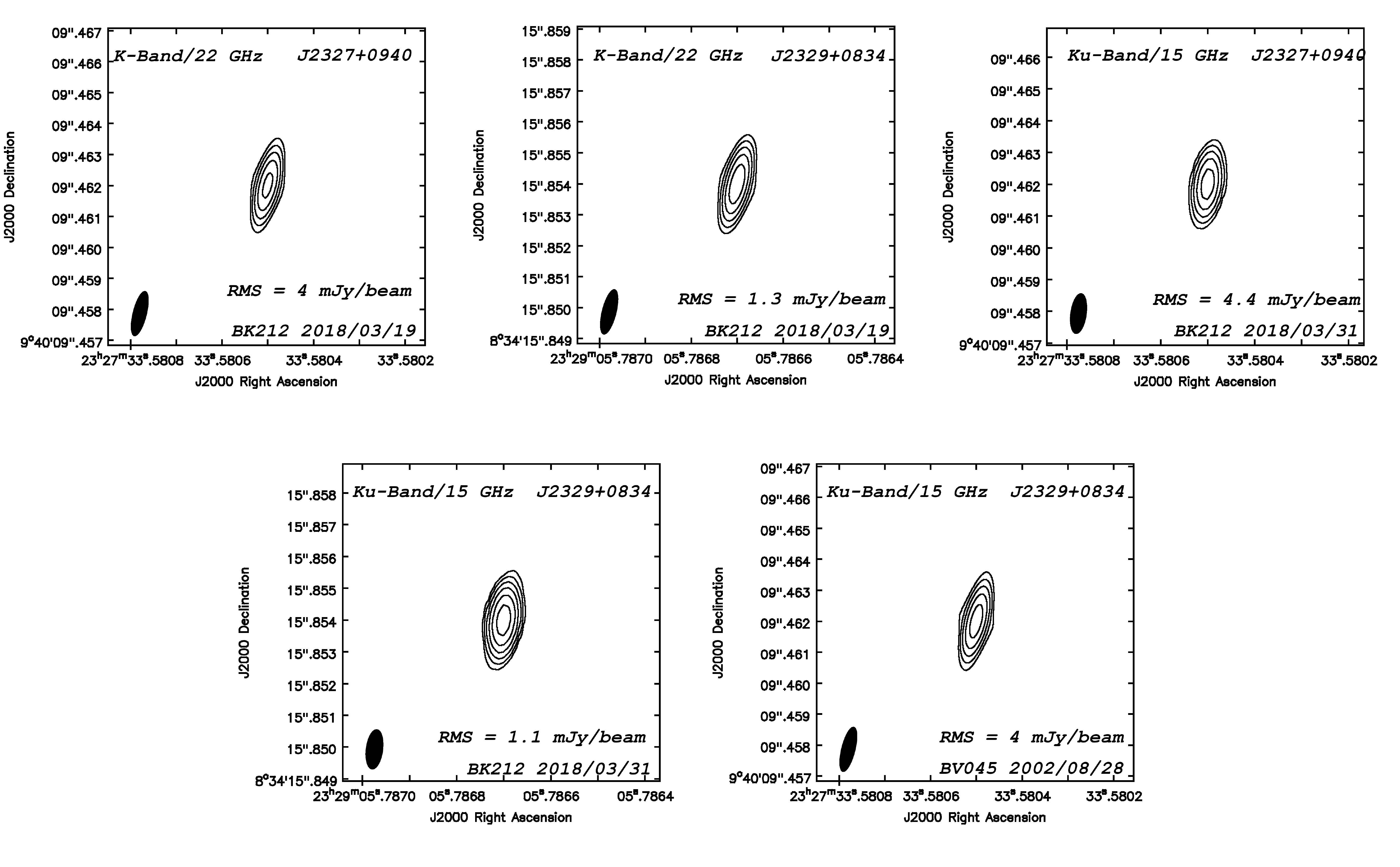}
	\end{center}
	\caption{Here we show the high-frequency VLBA Ku and K-band (15 and 22 GHz) images of our phase calibrators as contour plots.  The contours start at a base level of 5$\sigma$, and are spaced by factors of two thereafter.  The synthesized beams are shown as filled ellipses, and the image intensity noise level is specified by the RMS quoted in each contour plot.  The frequency and name of phase calibrator are given in the top of the plots, and project codes are also given along with the date of the observation in YYYY/MM/DD format.  The images are made folllowing the same CLEAN procedure defined in section~\ref{sec:imaging}.  No relevant structure which would impact the phase calibrtion is observed for any of our calibrators.}
	\label{fig:phasecal_images}
\end{figure}

\bibliography{ngc7674}{}
\bibliographystyle{aasjournal}

\end{document}